\documentclass[manuscript,screen]{acmart}
\AtBeginDocument{%
  }
\usepackage{subcaption}      
\usepackage{placeins}        
\setcopyright{acmlicensed}
\copyrightyear{2018}
\acmYear{2018}
\acmDOI{XXXXXXX.XXXXXXX}
\acmConference[TEI'26]{In Proceedings of the Eighteenth International Conference on Tangible, Embedded, and Embodied Interaction (TEI ’26)}{March 8-11, 2026}{CHICAGO,IL}
\acmISBN{978-1-4503-XXXX-X/2018/06}

\begin{document}

\title{Affective Translation: Material and Virtual Embodiments of Kinetic Textile Robots}


\author{Berfin Ataman}
\affiliation{%
  \institution{SMArchS Computation, MS Mechanical Engineering}
  \city{Cambridge}
  \state{MA}
  \country{USA}}
\email{bataman@mit.edu}

\author{Rodrigo Gallardo}
\affiliation{%
  \institution{SMArchS Computation, EECS}
  \city{Cambridge}
  \state{MA}
  \country{USA}}
\email{ragallar@mit.edu}

\author{Qilmeg}
\affiliation{%
  \institution{SMArchS Computation, EECS}
  \city{Cambridge}
  \state{MA}
  \country{USA}}
\email{qilmeg@mit.edu}

\renewcommand{\shortauthors}{ATAMAN et al.}


\begin{abstract}
This study presents a comparative framework for evaluating emotional engagement with textile soft robots and their augmented-reality (AR) counterparts. Four robotic sculptures were developed, each embodying nature-inspired dynamic behaviors such as breathing and gradual deformation. Using a between-subjects design, two independent groups, one experiencing the physical installations and one engaging with their virtual (AR) twins, follow identical protocols and complete the same self-assessment survey on affective and perceptual responses. This approach minimizes carryover and novelty effects while enabling a direct comparison of sensations such as calmness, curiosity, and discomfort across modalities. The analysis explores how motion, form, and material behavior shape emotional interpretation in physical versus digital contexts, informing the design of hybrid systems that evoke meaningful, emotionally legible interactions between humans, robots, and digital twins.
\end{abstract}



\received{25 October 2025}
\received[revised]{12 March 2009}
\received[accepted]{5 June 2009}


 \begin{teaserfigure}
   \centering
    \includegraphics[width=0.7\textwidth]{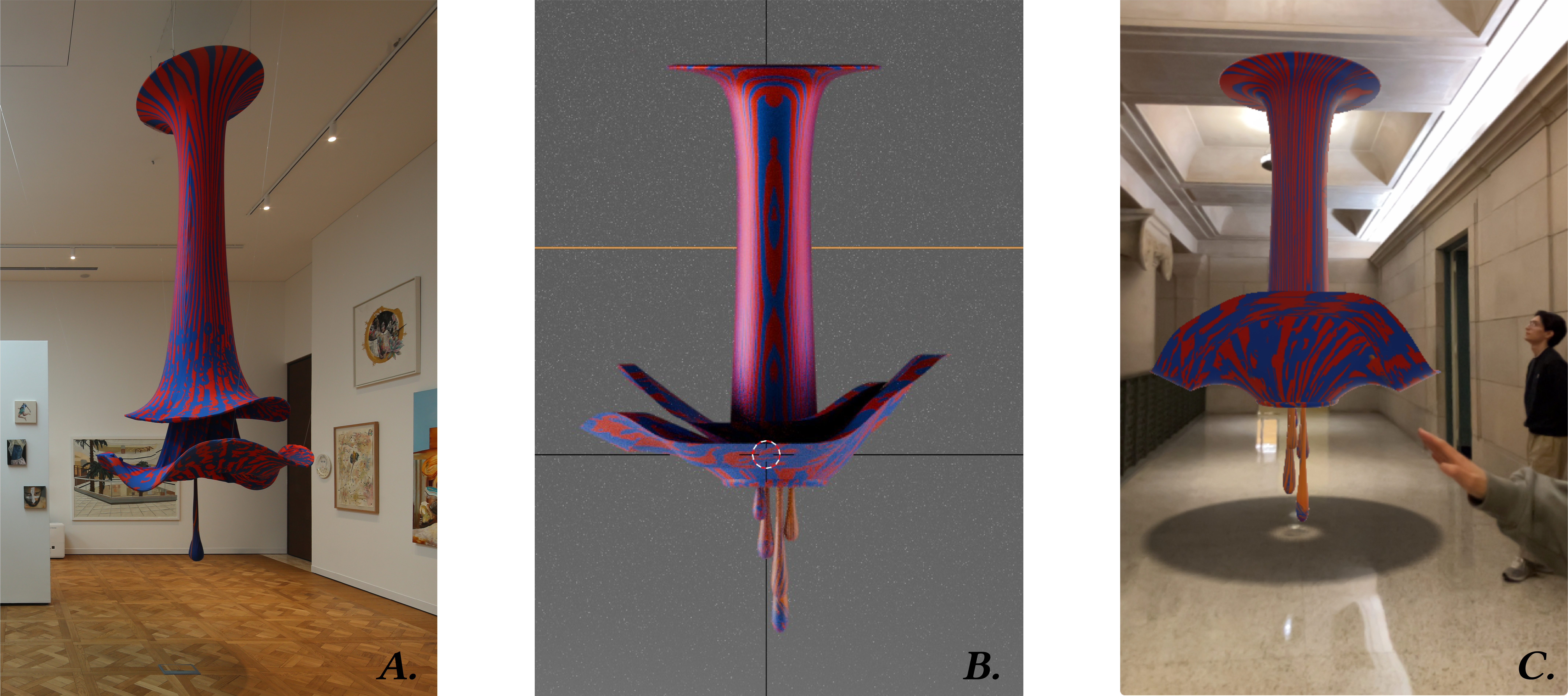}
    \caption{Overview of Affective Translation, showing (a) the physical textile soft-robot prototype illustrating scale and motion, (b) the textured digital model visualized in Blender, and (c) the augmented-reality twin replicating form and movement within a virtual environment.}
    \Description{figure description}
  \end{teaserfigure}

\maketitle

\section{Introduction}
Contemporary robotics often relies on recognizable representations. Humanoid and zoomorphic forms of faces, limbs, gestures, or animal mimicry are used to elicit empathy through familiarity. \textit{Affective Translation} proposes a different paradigm: textile soft robots inspired by visual and kinetic patterns in nature, such as breathing or wave-like deformation. This work builds on the concept of soft biomorphism \cite{christiansen_nature_2024} \cite{jorgensen_constructing_nodate}, translating natural dynamics into abstract form to evoke affective resonance.

We investigate how people emotionally engage with soft, nature-inspired motion. Parallel to the physical installations, we develop \textit{augmented reality (AR)} counterparts—digital twins of each textile robot—within an immersive, spatially aligned AR environment. These virtual versions reproduce the same motion but remove tactile feedback and material resistance, allowing us to study how emotional and perceptual responses shift when identical dynamics are experienced through digital mediation.

Soft robotics research increasingly moves beyond mimicry \cite{katzschmann_exploration_2018} toward material-driven adaptability. Research from Yale’s \textit{Faboratory} \cite{baines_robots_2024} demonstrates how robotic fabrics and variable-stiffness composites actuate in response to stimuli, redefining softness as a dynamic design property. These advances signal a shift from representation to \textit{expression through material behavior} \cite{raj_multimaterial_2025}. Our project extends this trajectory by asking not only how such systems move, but what emotions they evoke, and how these differ between physical and augmented experience.

By comparing responses between two independent groups—one engaging with the physical installations and another with their virtual counterparts—our goal is to map the \textit{emotional spectrum} that emerges when material motion is transposed into augmented space. This between-subjects approach reduces bias from carryover and learning effects, offering a clearer view of how different sensory conditions shape affective interpretation. The resulting cross-modal analysis informs the design of AR systems where physical and virtual elements coexist, highlighting how form, motion, and material behavior influence the emotions that guide human interaction.

\section{\textbf{Related Work}}
\subsection{{Representational Soft Robotics vs Material-Driven and Nature-Inspired Soft Robotics }}

Representational approaches elicit empathy through familiar forms and behaviors. Systems such as PARO and Joy for All engage users via gestures, faces, and speech to reduce anxiety and loneliness 

A material-driven trajectory treats softness as a morphological and technical property. Yale’s Faboratory shows robotic fabrics and variable-stiffness composites that morph, sense, and actuate across environments \cite{baines_robots_2024}. Complementary efforts include auxetic metamaterial skins with reversible, programmable deformation \cite{pu_robotic_2024} and woven soft active textiles using thin McKibben muscles \cite{li_variable_2025} \cite{guo2024encoded}. Reviews characterize this shift from mimicry to morphogenesis, where form, material, and motion are co-designed \cite{ambaye_soft_2024}.

Building on this foundation, Christiansen et al. explore biomorphic, nature-inspired abstraction \cite{christiansen_nature_2024}, and Jørgensen emphasizes sensation, ambiguity, and material responsiveness \cite{jorgensen_constructing_nodate}. Empirical findings show that breathing-inspired surfaces can reduce stress and increase calmness \cite{sabinson_every_2024}, and Sprout communicates internal states through deformation \cite{koike_sprout_2024}. Our work extends this lineage by comparing the same material behaviors across physical and augmented conditions.

\subsection{\textbf{Virtual Affect}}
Virtual embodiment studies show that users respond emotionally to abstract avatars when motion conveys living rhythms and intentionality, indicating that \textit{motion quality} can outweigh visual realism \cite{ambaye_soft_2024}. For example, immersive portals increase embodied storytelling in VR \cite{10.1145/3689050.3706013}, avatar customization affects affective engagement\cite{10.1145/3430524.3442446} , and users feel stronger connection to physical than AR objects even with matched geometry\cite{10.1145/3623509.3635255}. Yet direct comparisons of physical and virtual embodiments of the \textit{same} system are rare; “Sim-to-Real” typically targets control transfer rather than affective perception \cite{koike_sprout_2024}. We examine how identical motion patterns are interpreted when (a) physically embodied and (b) digitally rendered, probing how materiality, tactility, and resistance shape affective engagement.

\subsection{\textbf{Summary} }
\textit{Affective Translation} situates a shift from imitating living forms to exploring motion and material as expressive media. We contrast representational robots with material-driven systems and compare textile robots and their AR twins to reveal how identical motions elicit different emotional responses across embodiments, highlighting the roles of materiality, feedback, and motion quality.As interactions increasingly blend physical robots with augmented environments, this work highlights how embodiment shapes engagement in hybrid human–robot settings.

\section{Methodology}
\subsection{Overview}

This study examines how people emotionally and perceptually respond to textile soft robots across physical and virtual embodiments using a between-subjects design. Both systems share the same geometry, motion sequence, and temporal rhythm; however, the physical and virtual conditions are evaluated by two independent groups that complete identical procedures and instruments. This structure allows us to isolate modality effects while minimizing carryover, learning, and expectation biases that may arise in within-subject designs.

Each group experiences one condition only—either the physical installation or the virtual (AR) twin—under matched environmental and temporal conditions. Both groups engage with the work for the same duration and complete identical self-assessment surveys capturing their emotional and perceptual responses. This framework enables a direct comparison of affective qualities such as liveliness, calmness, curiosity, and connection across groups while ensuring that responses reflect the influence of embodiment and sensory modality rather than repeated exposure

\subsection{Participants and Study Design}

This study is currently ongoing and follows a between-subjects design involving two independent groups: a \textit{Physical Group} (\emph{n}=X) that engages exclusively with the physical textile soft robots, and a \textit{Virtual Group} (\emph{n}=Y) that interacts solely with their augmented-reality (AR) twins.

Each participant completes a single session consisting of a short orientation, an interaction period, and a post-experience self-assessment survey. Session durations, environmental lighting, and ambient sound levels are kept consistent across both conditions. No participant experiences more than one modality to minimize potential carryover, expectation, or novelty effects. 

Data collection and analysis are ongoing. Initial results for the Physical group below. Once both groups have completed participation, the responses will be compared to evaluate how embodiment and sensory modality influence affective and perceptual engagement with the textile soft robots and their virtual twins.

\subsection{ Building of the physical pieces} 

\subsubsection{Physical System Design} 

Various actuation systems for mechanics were used, such as spool systems, pneumatic chambers, and mechanical joints to evoke rhythms found in living organisms, such as breathing, pulsing, and swelling. Each robot contains a supporting structure that shapes the outer fabric form, combining lightweight materials such as wood, 3D-printed joints, and soft fillers.  These elements define the robots form and guide its deformation during motion. Each configuration was carefully tuned to allow the kinetic transformation to emerge smoothly, balancing rigidity with compliance.

\subsubsection{Surface Design }

Surface design plays a central role in shaping how motion is perceived. Prototyping in  white muslin allows one to read the form purely through its movement before introducing surface design. Once the behavior of a piece feels coherent, custom-printed fabrics are developed.  Across prototypes, three main strategies were tested for surface–form interaction: 

1. Independent patterning:  surface prints that disregard the object’s seams, causing the pattern to distort as the sculpture moves. This technique often conceals construction lines and produces a surprisingly lifelike quality. 

2. Contour-aligned patterning: patterns that trace the object’s sewn geometry, emphasizing edges and amplifying its volumetric presence. 

3. Interrupted patterning: hybrid surfaces where pattern fades intermittently into solid color, revealing underlying form only in fragments. 
The printed imagery comes from multiple sources: photographs of underwater organisms, traced and recolored natural(plant and animal) patterns, and vectorized designs for digital fabric prints. 

Each prototype’s fabrication required iterative calibration between material resistance and motion rhythm. Sewing techniques and actuator routing were adjusted to control motion direction and deformation. Every finished robot embodies a negotiation between mechanical logic and sensory affect.

\section{Actuation and Control Systems}
Each textile soft robot in the series uses a unique actuation system tailored to its physical design and behavior. Although all units share a digital infrastructure built on Adafruit Feather microcontrollers and remote data logging via Custom IO, their methods for movement spooling, inflation, and servo-driven tension differ, creating distinct temporal patterns. Randomized timing sequences are implemented across all robots to disrupt mechanical uniformity and mimic lifelike unpredictability.
\begin{figure}[H]
  \centering
  \begin{minipage}{0.48\linewidth}
    \centering
    \includegraphics[width=\linewidth]{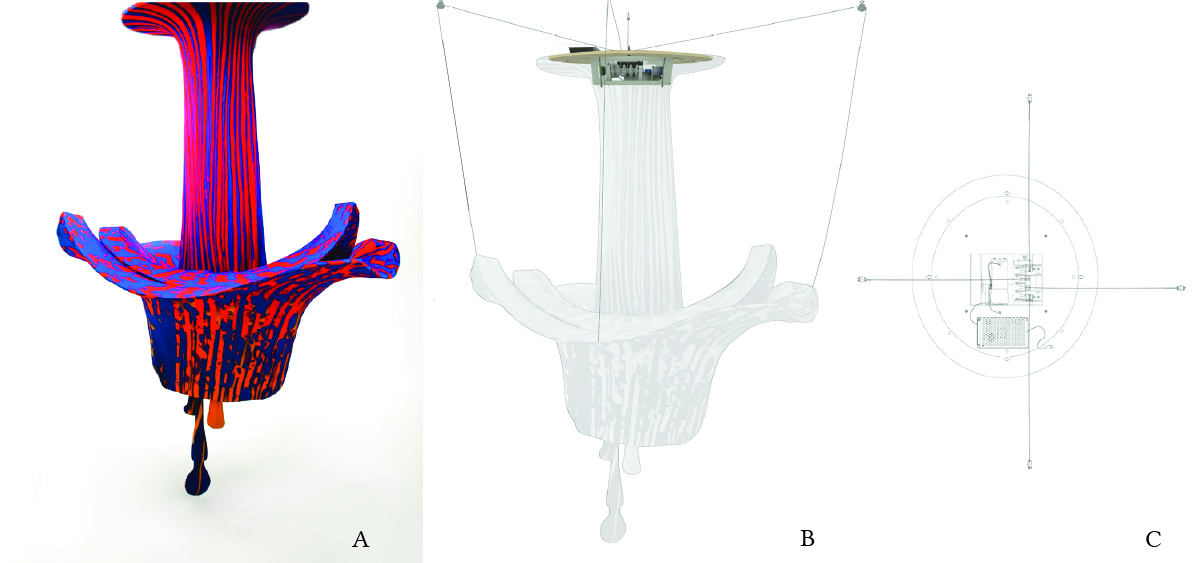}
    \caption*{Fig 2.  Waiting for the Dark Soft Robot. (a) Physical sculpture. (b) Placement of the system within the sculpture. (c) Birdseye view of the
motion system.}
\label{fig:waiting_dark}
  \end{minipage}\hfill
  \begin{minipage}{0.48\linewidth}
    \centering
    \includegraphics[width=\linewidth]{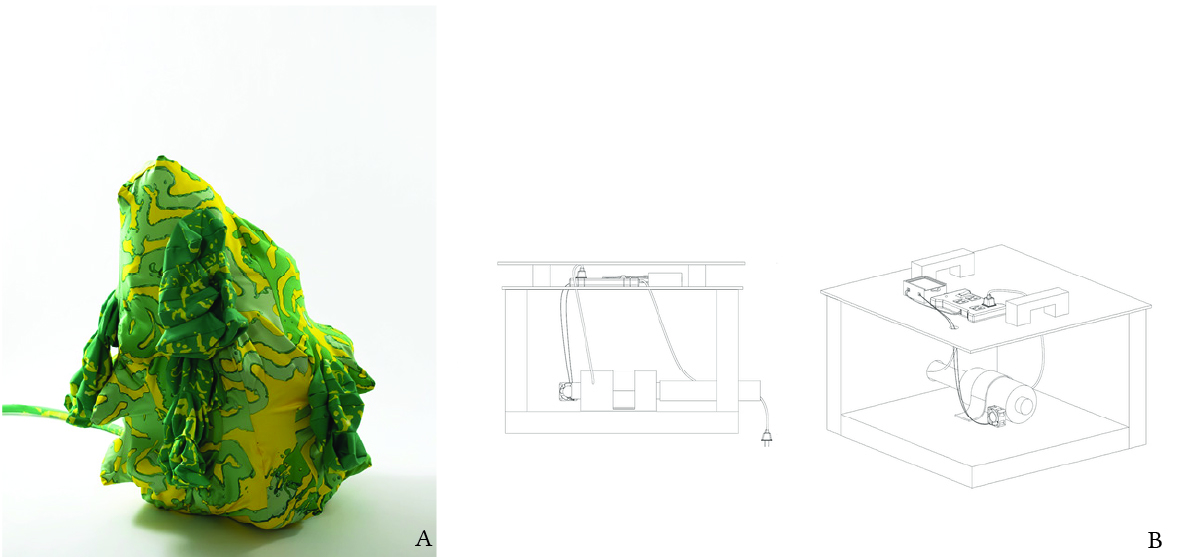}
    \caption*{Fig 3. Thirsty For air Soft Robot. (a) Pysical Sculpture. (b)Pneumonic System. }
  \end{minipage}
  \label{fig:comparison}
\end{figure}
\begin{figure}[H]
  \centering
  \begin{minipage}{0.48\linewidth}
    \centering
    \includegraphics[width=\linewidth]{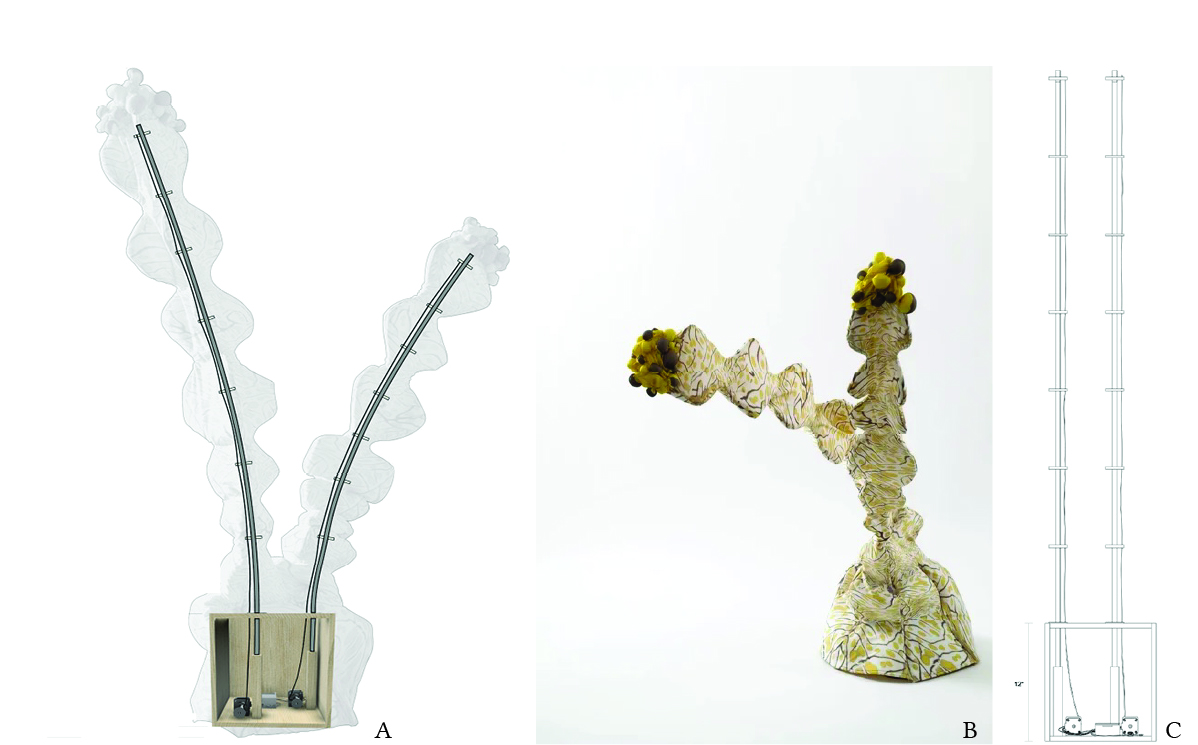}
    \caption*{Fig 4. Searching for the Sun Soft Robot. (a)Placement of the system within the robot shell. (b)Physical robot. (c) Supporting mechanical
motion system.}
  \end{minipage}
    \hspace{2em}
  \begin{minipage}{0.28\linewidth}
    \centering
\includegraphics[width=\linewidth]{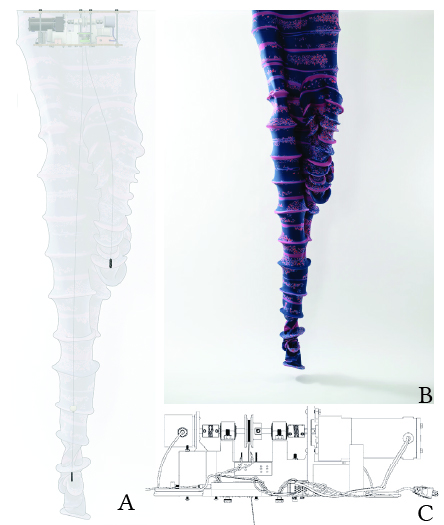}
    \caption*{Fig 5. Making Room. (a) Placement of the system within the robot's outer shell. (b) Physical robot. (c) Mechanical system that facilitates the motion.}
  \end{minipage}
  \label{fig:comparison}
\end{figure}

\setcounter{figure}{5} 

\subsection{ Waiting for the Dark}
\textit{Waiting for Dark} (Figure 2) moves with a single DC motor connected to four spools. Each spool winds a thread that lifts the bottom rim section of the fabric body upward, which then pulls up the smalled rim upwards with its movement creating a slow drawing motion reminiscent stretching.  An encoder attached to the motor provides feedback on rotational position, allowing the system to determine both the upper and lower limits of movement. These encoder readings are streamed, and the robot's motion can be monitored remotely in real time. The motor control firmware introduces randomized intervals within predefined bounds so that the lifting and lowering pattern never fully repeats. As a result, the piece appears always to change its motion, mechanically simple, yet perceptually complex.

\subsection{Thirsty for Air}
\textit{Thisrty for Air} (Figure 3) uses a compact air pump and a relay driven by a feather controller. The pump inflates the soft chamber for a set duration, holds briefly, and then deflates before repeating. The inflation–deflation cycle is randomized between two and three minutes of inflation, and randomized breaks in between inflation that lead to deflation. This  allows each breathing sequence to differ subtly from the last. The microcontroller records time-inflated data via the shared network, enabling feedback and remote control similar to the motorized systems. 

\subsection{Searching for the Sun}

\textit{Searching for the Sun's} (Figure 4), movement is produced by two servos housed within the robot’s internal skeleton. Each servo pulls nylon wires routed through the frame to create directional bending. The motion depends not only on servo rotation but also on the weight and drape of the textile skin, which resists and modulates the pull.

The servos are also controlled via the Adafruit Feather, sending live position and response data. The combined behavior, mimics the orientation gestures of plants tracking light, an association reinforced by the piece’s vertical posture and soft-edged silhouette.

\subsection{Making Room}

\textit{Making Room} (Figure 5) employs a dual-spool actuation system using one DC motor,  each spool controls an arm of the robot: a shorter and a longer fabric arm. Limit switches embedded at the top part of the robot calibrate the motor range, preventing overwinding and defining a consistent baseline. The system is equipped with encoders that transmit positional data, logging high-low status values in real time to the Adafruit IO interface. By alternating the winding and unwinding distances, Making Room creates a breathing rhythm that alternates between contraction and release. 

Across all systems, the networked Feather serves both as a control interface and sensing hub, merging soft robotics with the language of responsive environments. Each robot thus becomes a small, self-reporting organism—aware of its own movement, but never fully predictable.
This approach transforms actuation into a performative collaboration between code, mechanism, and material gravity, revealing motion not as automation but as embodied temperament.

\section{From Physical Prototypes to Virtual Twins}

The virtual twins replicate the four textile soft robots as fully animated VR models. Each twin is generated from high-fidelity 3D scans and CAD geometries modeled in Rhino and Blender, with fabric patterns UV-mapped to match the texture, color, and scale of the physical pieces. The animations are built in Unity using motion data recorded from the same microcontrollers that drive the physical systems, allowing both embodiments to share identical temporal signatures and amplitudes.

Within the virtual environment, participants in the \textit{Virtual Group} experience each piece through a head-mounted display that mirrors the spatial layout of the exhibition. The sculptures appear at full scale, occupying the same relative positions as their physical counterparts. Ambient spatial audio from pneumatic and tensile recordings enhances realism, while visual and auditory cues replace tactile feedback. By removing material resistance while preserving motion and scale, the digital twins reveal how sensory context—sight, sound, and spatial immersion—shapes perception of liveliness, empathy, and calmness.

As this work is in progress, data from the physical and virtual groups are being collected in parallel. Once both datasets are complete, cross-group analyses will compare affective and perceptual responses to identify how embodiment, material feedback, and sensory modality influence emotional engagement across conditions.

\begin{figure}[h]
  \centering
  \begin{minipage}{0.54\linewidth}
    \centering
    \includegraphics[width=\linewidth]{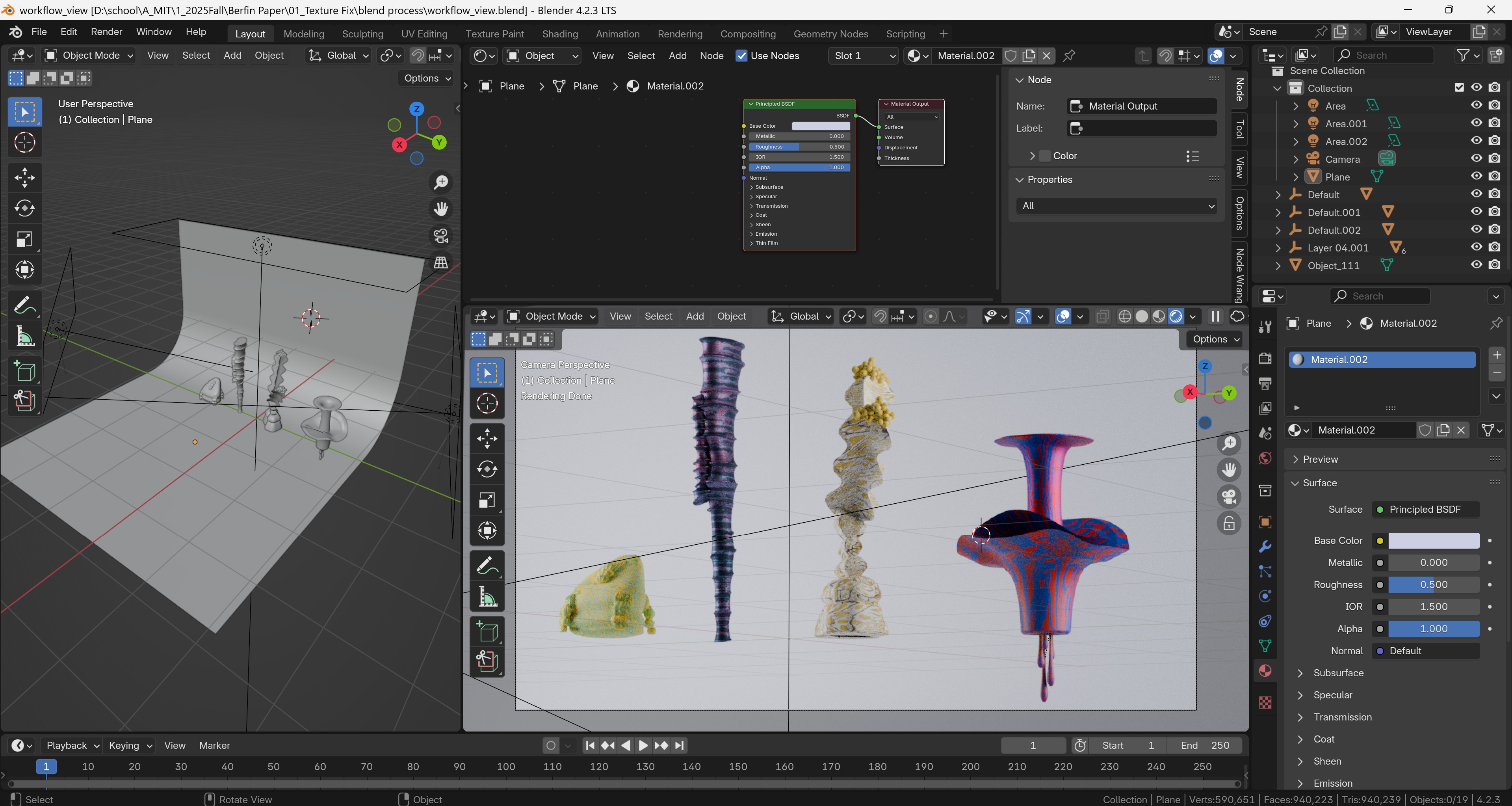}
    \caption{Texture mapping and visualization pipeline within Blender.}
    \Description{Stages of 3D model processing in Blender.}
  \end{minipage}\hfill
\end{figure}

\begin{figure}
    \centering
    \includegraphics[width=1\linewidth]{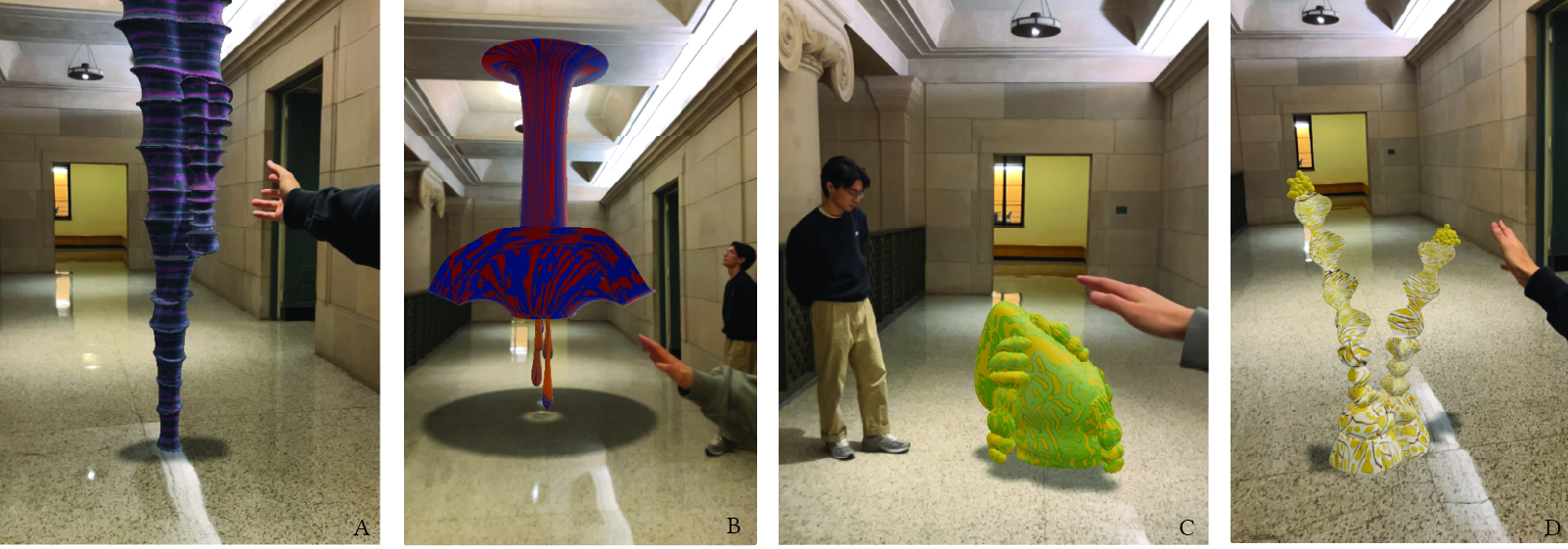}
    \caption{ All four robots visualized in the AR environment for interaction.}
    \label{fig:placeholder}
\end{figure}

\section{Data Collection}

Participants who experienced \textit{Waiting for the Dark}, \textit{Searching for the Sun}, \textit{Thirsty for Air}, and \textit{Making Room} through the physical installation were invited to complete a self-assessment survey. The survey was designed to capture perceptual, emotional, and tactile responses to each piece, focusing on how movement, color, and material presence influenced affect. The questionnaire was divided into thematic sections addressing participant context, emotional response, associative imagery, visual and material perception, and sense of presence and agency. 

To evaluate emotional and perceptual engagement across conditions, both the physical and virtual experiences used identical self-assessment surveys. After each session, participants rated affective qualities such as liveliness, calmness, curiosity, and connection. Open-ended responses captured descriptive language associated with motion and atmosphere, allowing for qualitative and quantitative comparison between material and digital embodiments.

Future iterations will integrate biometric sensing—heart rate variability, skin conductance, and pupil dilation—to complement self-reported affective data. This multimodal approach extends affective computing methods to embodied art installations, quantifying how emotional resonance translates between material and digital space.

Since the physical pieces were built prior to the AR versions, the data on virtual twins remains in progress. Once both parts of the experiment are complete, the collected responses will form a qualitative basis for analyzing how embodied interaction and sensory cues shape empathy, curiosity, and connection across physical and virtual embodiments of the work.

\begin{table}[h]
  \centering
  \caption{Average Participant Ratings for Waiting for the Dark (Physical Installation)}
  \label{tab:averages}
  \begin{tabular}{lc}
    \toprule
    \textbf{Question} & \textbf{Average Rating (1–5)} \\
    \midrule
    How pleasant was the experience? & 5.00 \\
    How engaging was it? & 4.70 \\
    How alive did it feel overall? & 4.40 \\
    How natural did its motion seem? & 4.40 \\
    How familiar did its form and color feel? & 4.22 \\
    \bottomrule
  \end{tabular}
\end{table}

\subsection{Initial Self Assessment Data Analysis}

\begin{figure}[H]
  \centering
  \includegraphics[width=.99\linewidth]{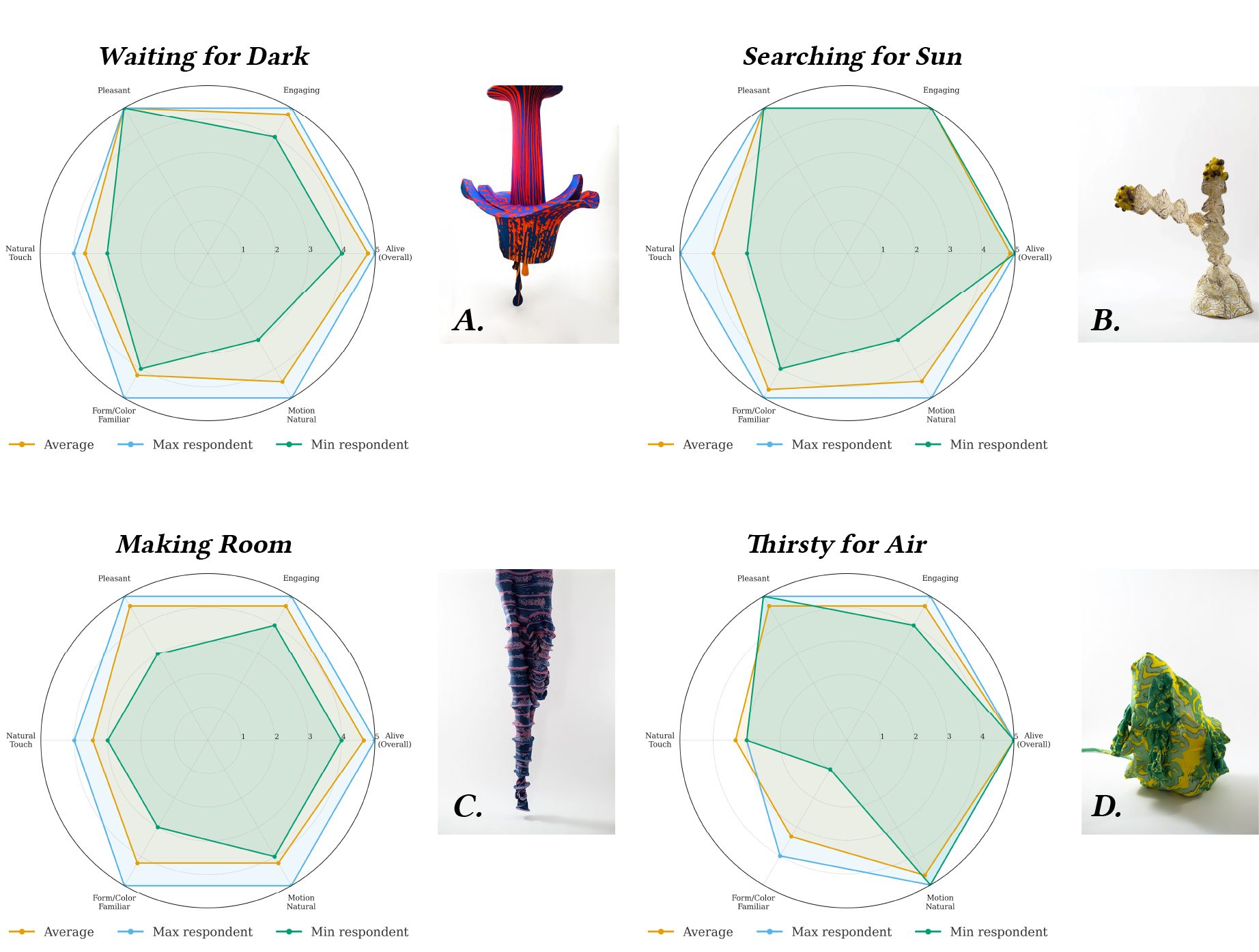}
  \caption{Mesh texturing and rendered output in Blender.}
  \Description{Two stages of 3D model processing in Blender.}
\end{figure}

With current data across participants, the general consensus reflects a highly positive and effectively engaging experience. Nearly all respondents rated the piece as pleasant, engaging, and “alive”, describing strong feelings of curiosity, calmness, and fascination. The installation’s motion was consistently perceived as natural and organic, though a few participants noted slight ambiguity regarding its familiarity, suggesting the work struck a balance between recognition and novelty.
Most respondents associated the motion with natural rhythms like breathing or wave movement, emphasizing a sense of gentle vitality rather than mechanical behavior. The radar plots reveal dense overlap in higher rating zones (4–5), indicating collective agreement that the experience felt emotionally resonant and lifelike, while still leaving room for individual interpretation. In short, the data points toward an installation that successfully evoked emotional and perceptual engagement through movement and material expression, fostering an atmosphere that felt alive, soothing, and contemplative without relying on literal representation.

\section*{GenAI Usage Disclosure}

Portions of this work were supported by generative AI tools in the writing, editing, and organization stages. Specifically, OpenAI’s ChatGPT (GPT-5, October 2025 version) was used to assist with language refinement, LaTeX formatting, and figure caption organization. All conceptual, analytical, and creative contributions, as well as code and experimental data, were developed entirely by the authors. The use of AI tools did not generate or analyze research data, nor did it contribute to the core results, figures, or findings presented in this paper.

\section{References}
\bibliographystyle{ACM-Reference-Format}
\bibliography{tei3}

\end{document}